\documentclass{aa}
\usepackage{natbib}
\usepackage{graphicx}
\usepackage{txfonts}
\usepackage[colorlinks=true,citecolor=blue]{hyperref}
\usepackage[dvipsnames]{xcolor}
\usepackage{amsmath}
\usepackage{amssymb}

\begin{document}

   \title{Asteroseismology of evolved stars to constrain the internal transport of angular momentum}

\subtitle{V. Efficiency of the transport on the red giant branch and in the red clump}
   \author{F.D. Moyano
          \inst{1}
          \and
          P. Eggenberger\inst{1}
          \and
          G. Meynet\inst{1}
          \and
          C. Gehan\inst{2,3}
          \and
          B. Mosser \inst{4}
          \and
          G. Buldgen \inst{1}
          \and
          S.J.A.J. Salmon \inst{1}
   }

   \institute{Observatoire de Gen\`eve, Universit\'e de Gen\`eve, 51 Ch. Pegasi, CH-1290 Versoix, Suisse \\
              \email{facundo.moyano@unige.ch}
              \and Max Planck Institut f\"ur Sonnensystemforschung, Justus-von-Liebig-Weg 3, 37077 G\"ottingen, Germany
              \and Instituto de Astrof\'isica e Ci\^encias do Espa\c{c}o, Universidade do Porto, CAUP, Rua das Estrelas, 4150-762 Porto, Portugal
              \and LESIA, Observatoire de Paris, Universit\'e PSL, CNRS, Sorbonne Universit\'e, Universit\'e de Paris, 92195 Meudon, France
             }

   \date{Received ----; accepted 05/05/22}


  \abstract
      {Thanks to asteroseismology, constraints on the core rotation rate are available for hundreds of low- and intermediate-mass stars in evolved phases.
        Current physical processes tested in stellar evolution models cannot reproduce the evolution of these core rotation rates.}
   {We investigate the efficiency of the internal angular momentum redistribution in red giants during the hydrogen shell and core-helium burning phases based on the asteroseismic determinations of their core rotation rates.}
   {We compute stellar evolution models with rotation and model the transport of angular momentum by the action of a sole dominant diffusive process parametrized by an additional viscosity in the equation of angular momentum transport.
   We constrain the values of this viscosity to match the mean core rotation rates of red giants and their behaviour with mass and evolution using asteroseismic indicators along the red giant branch and in the red clump.
      
}
   {For red giants in the hydrogen shell-burning phase the transport of angular momentum must be more efficient in more massive stars.
     The additional viscosity is found to vary by approximately two orders of magnitude in the mass range \mbox{M $\sim$ 1 - 2.5 M$_{\odot}$}.
     As stars evolve along the red giant branch, the efficiency of the internal transport of angular momentum must increase for low-mass stars (M $\lesssim$ 2 M$_{\odot}$) and remain approximately constant for slightly higher masses (2.0 M$_{\odot}$ $\lesssim$ M $\lesssim$ 2.5 M$_{\odot}$).
     In red-clump stars, the additional viscosities must be an order of magnitude higher than in younger red giants of similar mass during the hydrogen shell-burning phase.
}
   {In combination with previous efforts, we obtain a clear picture of how the physical processes acting in stellar interiors should redistribute angular momentum from the end of the main sequence until the core-helium burning phase for low- and intermediate-mass stars to satisfy the asteroseismic constraints.
    }

   \keywords{stars: rotation --
             stars: interiors  --
             stars: evolution --
             methods: numerical  }

   \maketitle


   \section{Introduction}
   The impact of rotation on the structure of stars and their evolution has been addressed since the early time of stellar physics \citep[see e.g.][]{VZ1924,Eddington1929, Sweet1950, Opik1951, Mestel1988}. Grids of evolutionary rotating models have then been computed \citep[see e.g.][]{EndalSofias1978, Pins1989, Deupree1990, Fliegner1996,Heger2000,MM2000,Palacios2006, ekstrom12, Choi2017, Lim2018} based on the above first results and also 
   on more recent developments \citep[as e.g.][]{Zahn1992,spruit02}. 
   Despite these long-term efforts, the transport of angular momentum (AM) in stellar interiors remains an open question.
   Indeed, our current picture of hydrodynamical transport of AM is insufficient to explain the rotation periods of compact objects \citep{suijs08} and does not reproduce the internal angular velocity distribution deduced from helio- and astero-seismic analyses \citep[see the discussion in][and the next paragraph]{eggenberger19c}.

   The long and continuous data sets obtained by space-borne missions, such as CoRoT \citep{baglin09}, \textit{Kepler} \citep{borucki10}, TESS \citep{ricker15}, and in the future PLATO \citep{rauer14}, led to the detection of mixed oscillation modes in evolved stars, which enabled to reveal their internal rotation.
   Interestingly, mixed oscillation modes are simultaneously sensitive to the physical properties in the central and external layers of a star. In particular, it was possible to gain information about the core rotation rates of subgiant and red giant stars at different evolutionary stages \citep{bec12,deh12, mosser12,deheuvels14,deh15,dim16,tri17,gehan18,dimauro18,tay19,deheuvels20}.
   These data sets allow one to investigate in a statistical way the behaviour of AM redistribution in stellar interiors during their evolution in the subgiant and red giant phases.
   Nowadays, we know that stellar models including only hydrodynamical processes fail to reproduce the angular velocity of stellar cores in different evolutionary phases \citep[e.g.][]{eggenberger12,marques13,cei13}.
   In particular, for stars on the lower red giant branch (RGB), the cores are rotating too fast by up to three orders of magnitude compared to asteroseismic constraints.
   This indicates that at least one efficient additional AM transport process must be at work in the interior of evolved stars, similarly to what is found for the Sun \citep{eggenberger05,eggenberger19b} or for the rotation rates of compact objects \citep{suijs08}.

   Several physical processes are candidates for this missing AM transport mechanism.
   Internal magnetic fields can transport AM efficiently in radiative regions and slow-down the stellar cores.
   A first possibility relies on fossil magnetic fields to ensure uniform rotation in stellar radiative zones and the assumption of radial differential rotation in the convective envelopes of evolved stars to account for the core rotation rates observed for these stars \citep{kis15, tak21}. This hypothesis is however disfavoured by asteroseismic measurements of red giants, which are not compatible with a uniform rotation in the radiative interior of these stars \citep{kli17, fellay21}. Magnetic instabilities can also lead to an efficient AM transport in stellar radiative zones \citep{spruit99,spruit02}.
   Although they may be easily triggered above a certain degree of differential rotation, they are strongly inhibited by chemical gradients.
    Consequently, rotating models based on the Tayler-Spruit dynamo \citep{spruit02} are found to predict an internal coupling for red giants that is not efficient enough to reproduce their observed core rotation rates \citep{can14,denhartogh19}.
   A revised formulation for this process leading to a more efficient AM transport than the Tayler-Spruit dynamo has then been proposed \citep{fuller19}. Lower core rotation rates are then obtained for evolved stars, which are in better global agreement with observed values, but asteroseismic measurements of subgiant, red giant and secondary clump stars cannot be correctly reproduced \citep{eggenberger19c,den20}. Internal gravity waves may also transport AM in an efficient way.
   These waves can be excited by turbulent motions in convective zones \citep[e.g.][]{press81}, as well as by penetration of convective plumes \citep[e.g.][]{pincon16,pincon17}. Gravity waves (especially the ones excited by penetrative convection) could play a role in shaping the rotation profile during the subgiant phase, but seem to be unable to influence the rotation in the central layers of red giant stars \citep[][]{pincon17}.
   Finally, AM transport by mixed oscillation modes could be a promising candidate for the upper part of the red giant branch \citep{belkacem15}.
   Although we highlight that neither internal gravity waves nor mixed modes can extract AM efficiently from the cores of stars close to the RGB base, where they might spin up considerably.
   The physical process responsible for the AM transport in this rapid phase is still unknown.
   
   We thus see that we are still far from having a physical explanation to the internal rotation observed in evolved stars. It is then valuable to use asteroseismic measurements to characterise at best the properties of this unknown transport process to try to reveal its physical nature. 
   We do this in a similar way as in previous works \citep[see][]{eggenberger12,spada16,eggenberger17,eggenberger19a,denhartogh19}, although in this paper we focus solely on the evolution of the core rotation rate and take advantage of a larger number of stars with well characterized asteroseismic properties (see Sect. \ref{data}).
     This in turn allows us to draw conclusions on the role of the stellar mass and evolution along the red giant phase.
     We also compute low-mass stellar models in the core-helium burning phase, hence complementing the efforts done for the more massive secondary-clump stars \citep[e.g.][]{tayar13,tayar18,tay19,denhartogh19}.
   We parametrize the efficiency of the unknown transport process by a constant additional viscosity ($\nu_{\rm add}$) in the equation of AM transport. Of course, we do not expect the undetermined AM transport process(es) to be effectively described by this simple assumption of a constant viscosity in stellar radiative zones, but this approach enables us to quantify the efficiency with which they must operate to satisfy the empirical constraints.
   More precisely, it enables to probe how the efficiency of this additional transport process has to vary with mass and time in different evolutionary phases.
   This can put constraints on the nature of the physical processes acting in stable stratified regions.
   The mean viscosities derived in this work may also serve as benchmark values for ongoing theoretical developments and numerical multi-dimensional simulations, as for instance the AM transport by the azimuthal magnetorotational instability \citep[AMRI;][]{rud14} or by the Goldreich-Schubert-Fricke instability \citep[GSF;][]{barker19,barker20}. 

   Asteroseismic constraints indicate that, in subgiants, the AM redistribution must occur very efficiently just after the main sequence (MS) and then gradually weaken towards the RGB \citep{eggenberger19b,deheuvels20}.
   AM redistribution has also been found to be more efficient as stars ascend the RGB with an efficiency that increases with the stellar mass \citep{eggenberger17}.
   The change of efficiency with evolution between the subgiant and the red-giant phase might point to different physical processes acting in these two phases or, if the processes are the same, to significantly different consequences due to different physical conditions in stellar interiors.
   For instance, the timescale for the crossing of the Hertzsprung gap and the ascent of the RGB are governed by different processes.
   While this crossing is linked to the contraction of the core and the expansion of the envelope on a rapid timescale, the ascent of the RGB occurs on a longer nuclear timescale.
   Indeed in this last case, the evolution is due to the growing helium core resulting from the activity of the H-burning shell.
   The transition between these two phases is thus a key point to understand.

   The previously mentioned study of \citet{eggenberger17} was based on two well characterized targets so that a confirmation of these trends with larger samples is still lacking.
   Here we aim to fill this gap using a sample of hundreds of red giants obtained by \citet{gehan18} with determined rotational splittings of dipole mixed modes and asteroseismic quantities.
   This enables us to better characterize their evolutionary stage .
   We finally complement our study with data of red-clump stars using the sample of \citet{mosser12}.

   In Sect. \ref{methods} we describe the tools used and the input physics of the stellar models. In Sect. \ref{results} we present the models computed and the additional viscosities inferred to discuss their implications and global behaviour along evolution in Sect. \ref{discussion}. We summarize our main results in Sect. \ref{conclusion}.
\section{Stellar models and data}
\label{methods}
\subsection{Input physics: red giants}
We compute stellar evolution models of red giants, starting from the zero age main sequence (ZAMS), with the Geneva stellar evolution code \citep[{\fontfamily{qcr}\selectfont GENEC};][]{eggenberger08}.
We refer the reader to the mentioned work for details on the input physics and numerical treatment.
Here we give only a brief explanation of the most relevant parameters used to compute the models presented in this work and in particular for the modelling of AM transport in stellar interiors.
Rotation is treated assuming shellular rotation \citep{Zahn1992}.
The equation to follow the evolution of the AM is treated in an advecto-diffusive way (see Eq. \ref{eq_amt_genec}).
In convective zones we assume solid body rotation.
In radiative zones, the following equation:
\begin{equation}
  \rho \frac{{\rm d}}{{\rm d}t} \left( r^{2}\Omega \right)_{M_r}
  =  \frac{1}{5r^{2}}\frac{\partial }{\partial r} \left(\rho r^{4}\Omega
  U(r)\right)
  + \frac{1}{r^{2}}\frac{\partial }{\partial r}\left(\rho D r^{4}
  \frac{\partial \Omega}{\partial r} \right) \
\label{eq_amt_genec}
\end{equation}
is solved, where \textit{r} and $\rho$ stand for radius and mean density on an isobar,  $\Omega$ is the horizontally averaged angular velocity, \textit{U(r)} expresses the radial dependence of the vertical component of the meridional circulation velocity, and \textit{D} is the diffusion coefficient.
The diffusive term of this equation (the second one on the right side) is responsible for decreasing the differential rotation in radiative regions as it acts on the shear, denoted here by $\partial \Omega/ \partial r$.
The diffusion coefficient \textit{D} takes into account the diffusive processes as a linear sum of each of them.
For the diffusion of AM we include the shear instability \citep[$D_{\rm shear}$;][]{maeder97} and a constant additional viscosity ($\nu_{\rm add}$), hence $D=D_{\rm shear}+\nu_{\rm add}$.

In all of our models we adopt a solar metallicity with the solar chemical mixture given by \citet{asplund09}.
We also use a solar-calibrated value for the mixing-length parameter. 
The remaining physical ingredients, such as overshooting and mass-loss, among others, are the same as in \citet{ekstrom12}.

\subsection{Input physics: core-helium burning stars}
To explore the transition towards the core-helium burning phase of low-mass stars, we use the {\fontfamily{qcr}\selectfont MESA} stellar evolution code \citep{paxton11,paxton13,paxton15,paxton19}, version 15140\footnote{The input files and extensions necessary to reproduce our models are available at \url{https://zenodo.org/record/6408548}}. 
The use of the {\fontfamily{qcr}\selectfont MESA} code enables us to study the internal rotation of red clump stars (low-mass stars in the core-helium burning phase), since the {\fontfamily{qcr}\selectfont GENEC} code does not routinely follow the rotational evolution of stars after the helium flash. Moreover, using two evolutionary codes with different implementations of rotational effects enables us to check that the results obtained about the AM transport efficiency in stars ascending the RGB are not too sensitive to the specific code used.

Rotation in {\fontfamily{qcr}\selectfont MESA} is treated following \citet{Heger2000}, and we refer the reader to that work for further details.
Although the treatment of rotation in {\fontfamily{qcr}\selectfont MESA} is purely diffusive, while it is an advecto-diffusive process as treated in {\fontfamily{qcr}\selectfont GENEC}, we verified that the results and conclusions derived in this work are independent of the code used.
This is due to the fact that AM transport by meridional currents and the shear instability is negligible in the type of evolved stars that we study in this work.
 The equation used in {\fontfamily{qcr}\selectfont MESA} to follow the evolution of the AM transport is:
\begin{equation}
\left( \frac{\partial \Omega}{\partial t}\right)_m=\frac{1}{j}\left( \frac{\partial}{\partial m}\right)_t \left[ (4\pi r^2 \rho)^2jD \left( \frac{\partial \Omega}{\partial m}\right) \right]-\frac{2\Omega}{r}\left(\frac{\partial r}{\partial t}\right)_m\left( \frac{1}{2}\frac{\rm{d ln}\textit{j}}{\rm{d ln}\textit{r}}\right)
\label{eq_amt_mesa}
\end{equation}
where \textit{j} is the specific angular momentum.
The rest of the variables are defined as in Eq. (\ref{eq_amt_genec}).
The additional viscosity $\nu_{\rm add}$ is added linearly in the diffusion coefficient \textit{D} in the first term.
For consistency with our {\fontfamily{qcr}\selectfont GENEC} models, we only account for transport of AM by the shear instability and meridional circulation (in addition to $\nu_{\rm add}$) in radiative regions.

\subsection{Asteroseismic constraints}
\label{data}
To study the redistribution of AM at different evolutionary phases and for different mass ranges we compiled a data set with constraints on the core rotation rate for low- and intermediate-mass stars.
The core and surface rotation rate of eight subgiants was obtained by \citet{deheuvels14,deheuvels20}.
As for red giants, 875 red giants were studied in a systematic way and their rotational splittings made available by \citet{gehan18}.
The mass of these stars, as estimated by asteroseismic scaling relations \citep[e.g.][]{kallinger10}, is in the range M $\sim 1 $ -- 2.5 M$_{\odot}$.
  \begin{figure}[htb!]
     \resizebox{\hsize}{!}{\includegraphics{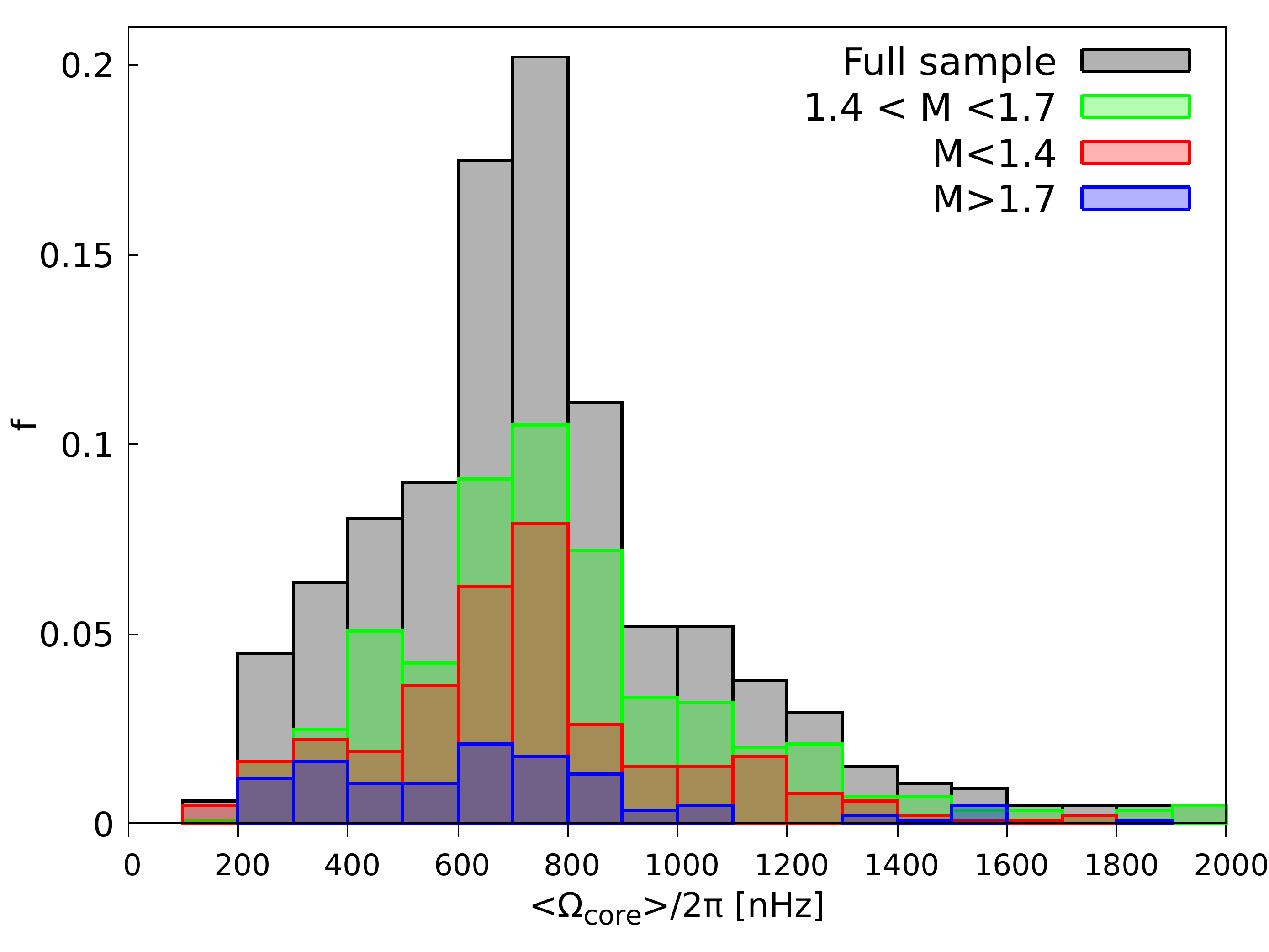}}
     \caption{Fraction of stars as a function of the mean core rotation rate for the full sample of hydrogen-shell burning stars from \citet{gehan18}, and for different stellar mass ranges as indicated in the figure in units of solar masses.}
     \label{histogram_omega}
   \end{figure}
We also benefit from the asteroseismic quantities derived in the mentioned work, namely the large frequency separation ($\Delta\nu$), the period spacing ($\Delta\Pi_{1}$) and the frequency of maximum oscillation signal ($\nu_{\rm max}$).
We used the rotational splittings of the most g-dominated modes, as given by \citet{gehan18} to estimate the core rotation rate of all stars in the sample.
This can be done under the assumption that gravity-dominated dipole mixed modes are mainly sensitive to the core.
However, the rotational splittings in the lower part of the red giant branch (before the bump) could be affected by the envelope.
To account for this contamination, a correction factor can be introduced with help of the asteroseismic quantities mentioned above \citep{mosser12}.
In this way, the mean core rotation rate can be obtained from the rotational splittings as \citep{gou13}:
\begin{equation}
  \langle\Omega_{\rm core}\rangle = 4\pi\eta \delta\nu_{\rm rot, core}
\end{equation}
where $\delta\nu_{\rm rot, core}$ is the measured core rotational splitting, and $\eta$ is the correction factor mentioned above.
This correction factor can be computed as
\begin{equation}
  \eta \simeq 1+\frac{\gamma}{\mathcal{N}}
  \end{equation}
where $\gamma \simeq 0.65$ \citep{mosser12} and $\mathcal{N}$ is the mixed mode density (see Eq. \ref{eq-mixmod} below).
This correction factor $\eta$ is slightly larger than unity for stars close to the RGB base and decreases towards unity as they ascend it.
Using this correction factor decreases the margin of error on the mean core rotation rate obtained from g-dominated mixed modes instead of pure gravity ones. Quantifying the difference in the retrieved mean core rotation rate when using g-dominated mixed modes instead of pure gravity modes would require the use of rotational kernels derived from stellar structure models, and the assumption of a rotation profile. This is beyond the scope of this work. Nevertheless the envelope contribution to the splitting of g-dominated modes is negligible \citep{mosser12}, hence we do not expect important differences.
Additionally, the uncertainties from \citet{gehan18} on the measured rotational splittings can be translated into uncertainties on the estimated core rotation rate. However, they are on the order of $\sim 10$ nHz which leads to a  maximum uncertainty of $\sim 30$ nHz on the mean core rotation rate for our hydrogen-shell burning stars. Thus, for clarity we decide not to include the associated error bars in any of our figures since in many cases they would be even smaller than the symbol size.

  The resulting distribution of core rotation rates for different mass ranges is shown in Fig. \ref{histogram_omega}. We do not intend to re-analyze the distribution seen for these stars, since it was already done by \citet{gehan18}. Nevertheless, it is worth recalling that most of these stars have core rotation rates in the range $\langle\Omega_{\rm core}\rangle/2\pi \in [600, 800] $ nHz, a mean mass of $\sim 1.5 M_{\odot}$, and their core rotation rates do not show any dependence with mass nor evolution \citep[see Figs. 10 \& 11 from][]{gehan18}.

As suggested by \citet{gehan18}, we use the mixed mode density ($\mathcal{N}$) as a proxy for the specific evolutionary phase along the RGB.
This physical quantity is equal to the number of gravity modes, within a frequency range whose width is equal to the large frequency separation.
It is defined as
\begin{equation}
  \label{eq-mixmod}
  \mathcal{N} = \frac{\Delta\nu}{\Delta\Pi_{1} \nu^{2}_{\rm max}}
\end{equation}
To compare it with our model results, we compute the asymptotic large frequency separation. The $\nu_{\rm max}$ frequency is obtained reversely using global asteroseismic scaling relations \citep[see e.g. ][]{kje95,kallinger10}.
 As for the period spacing, we compute its asymptotic estimate from the structure of each model as
\begin{equation}
\Delta\Pi_{1} =\frac{2\pi^{2}}{\sqrt2} \left(\int_{0}^{r_{\rm g}} {\rm{N_{\rm BV}}} \frac{dr}{r}\right)^{-1}
\end{equation}
where N$_{\rm BV}$ is the Brunt-V\"ais\"al\"a frequency and \textit{r} is the radial coordinate. The integral is done over the g-mode cavity, with r$_{\rm g}$ the radius of the upper limit of the cavity.
  \begin{figure}[htb!]
     \resizebox{\hsize}{!}{\includegraphics{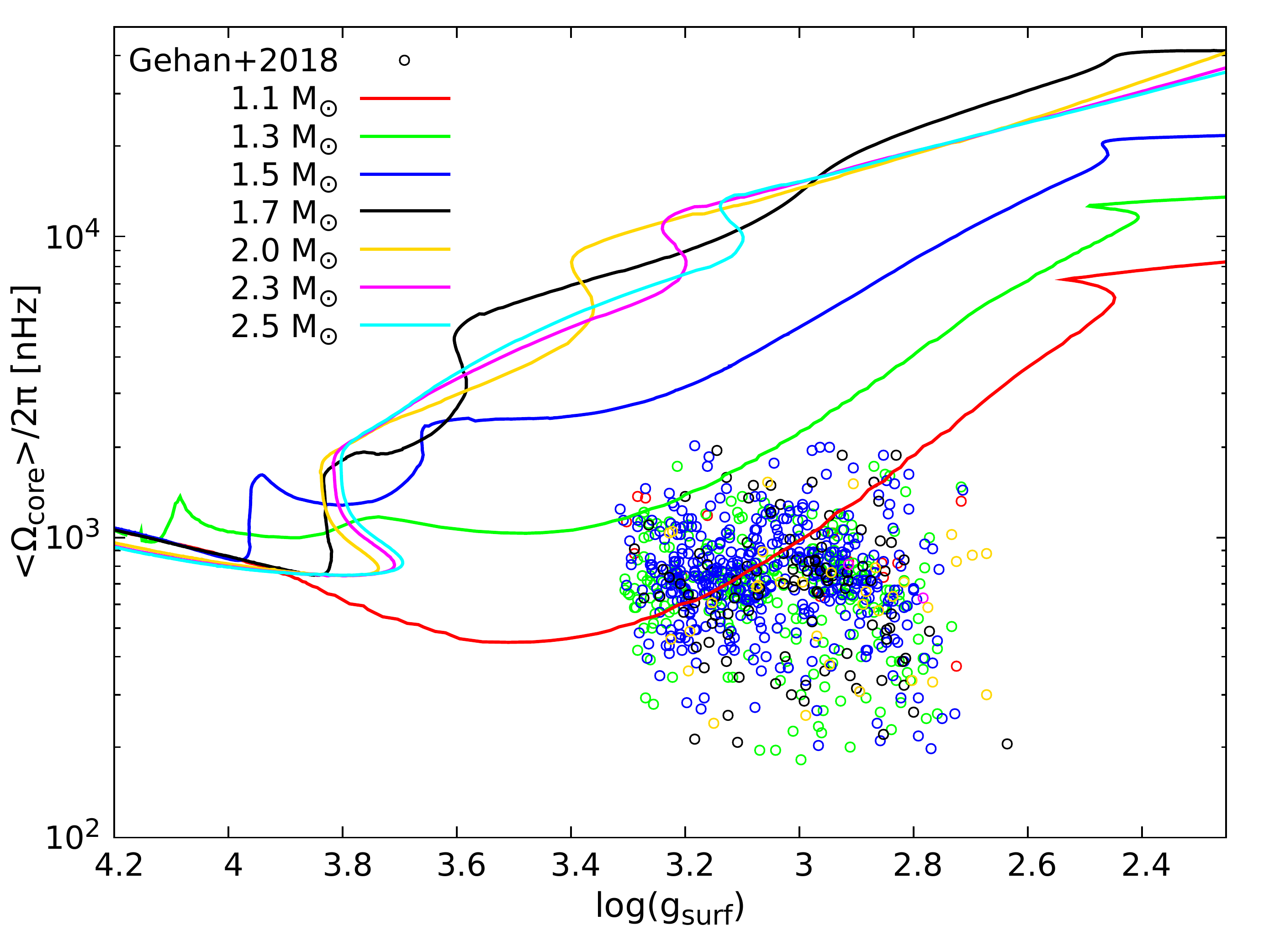}}
     \caption{Evolution of the core rotation rate as a function of the surface gravity. The data points correspond to the red giants in the hydrogen-shell burning phase analysed by \citet{gehan18} and are colour-coded by their mass as for the models. Each model has a different initial mass as indicated in the figure. An additional viscosity of $\nu_{\rm add}=8 \times 10^{3}$ cm$^{2}$/s was used in all models.}
     \label{omegac_gsurf}
   \end{figure}
The core rotation rate inferred with asteroseismic techniques is an average value of the angular velocity near the core, because it is related to the regions tested by gravity modes; it is not the angular velocity of the centre.
Hence, to compare directly our results with the constraints available from asteroseismic techniques we compute the mean core rotation rate $\langle\Omega_{\rm core}\rangle$ from our models as \citep{gou13}
\begin{equation}
  \langle\Omega_{\rm core}\rangle= \frac{\int_{0}^{r_{\rm g}}{\Omega \rm N_{\rm BV} dr/r}}{\int_{0}^{r_{\rm g}}{\rm N_{\rm BV} dr/r}}
\end{equation}
where $\Omega$ is the angular velocity and \textit{r} the radial coordinate. This integral is computed over the g-mode cavity, and represents the angular velocity in the near-core region as sensed by gravity modes.

Finally, for low-mass stars in the core-helium burning phase, we use the constraints on the mean core rotation rate given by \citet{mosser12}.
In this case, we only use the mean core rotation rate, and the mass and radius estimated via scaling relations. We do not use the mixed mode density as an evolution indicator for core-helium burning stars in this work.

\section{Efficiency of angular momentum redistribution}
\label{results}

In this section we present the results on the efficiency needed for the AM transport to satisfy the asteroseismic constraints presented in Sect. \ref{data}.
 We first present the role of the stellar mass on the efficiency of the angular momentum redistribution for red giants in the hydrogen-shell burning phase (Sect. \ref{sect_rgbmass}).
  Then we show how this efficiency should vary with evolution along the lower-RGB (Sect. \ref{sect_rgbevol}).
  We finally present a similar analysis for low-mass (M $\lesssim 2$ M$_{\odot}$) core-helium burning stars (Sect. \ref{sect_redclump}).

\subsection{Mass dependence in red giants}
\label{sect_rgbmass}
To investigate the mass-dependence of the additional viscosity needed to satisfy the asteroseismic constraints, we computed a grid of models with different initial masses and additional viscosities.
We computed models in the mass range of \mbox{1.1 to 2.5 M$_{\odot}$}, which covers the whole mass range in the data set presented in Sect. \ref{data}, according to scaling relations.
In this series of models we start the models from the ZAMS and adopt an initial period of $P=10$ days (see Sect. \ref{uncertainties} for a discussion on the impact of this choice).
We chose this initial period since it roughly reproduces the surface rotation rates of subgiants for which a detailed asterosesimic characterization of their internal rotation was obtained by \citet{deheuvels14}.
For each model with a different initial mass, we adopted a range of values for $\nu_{\rm add}$ suitable to reproduce the bulk of the data.
To refine the estimates on $\nu_{\rm add}$ for each mass value we ultimately interpolate between tracks with same initial mass and different $\nu_{\rm add}$ when the difference in $\nu_{\rm add}$ between adjacent models is small enough, typically less than 10\%.
  \begin{figure}[htb!]
     \resizebox{\hsize}{!}{\includegraphics{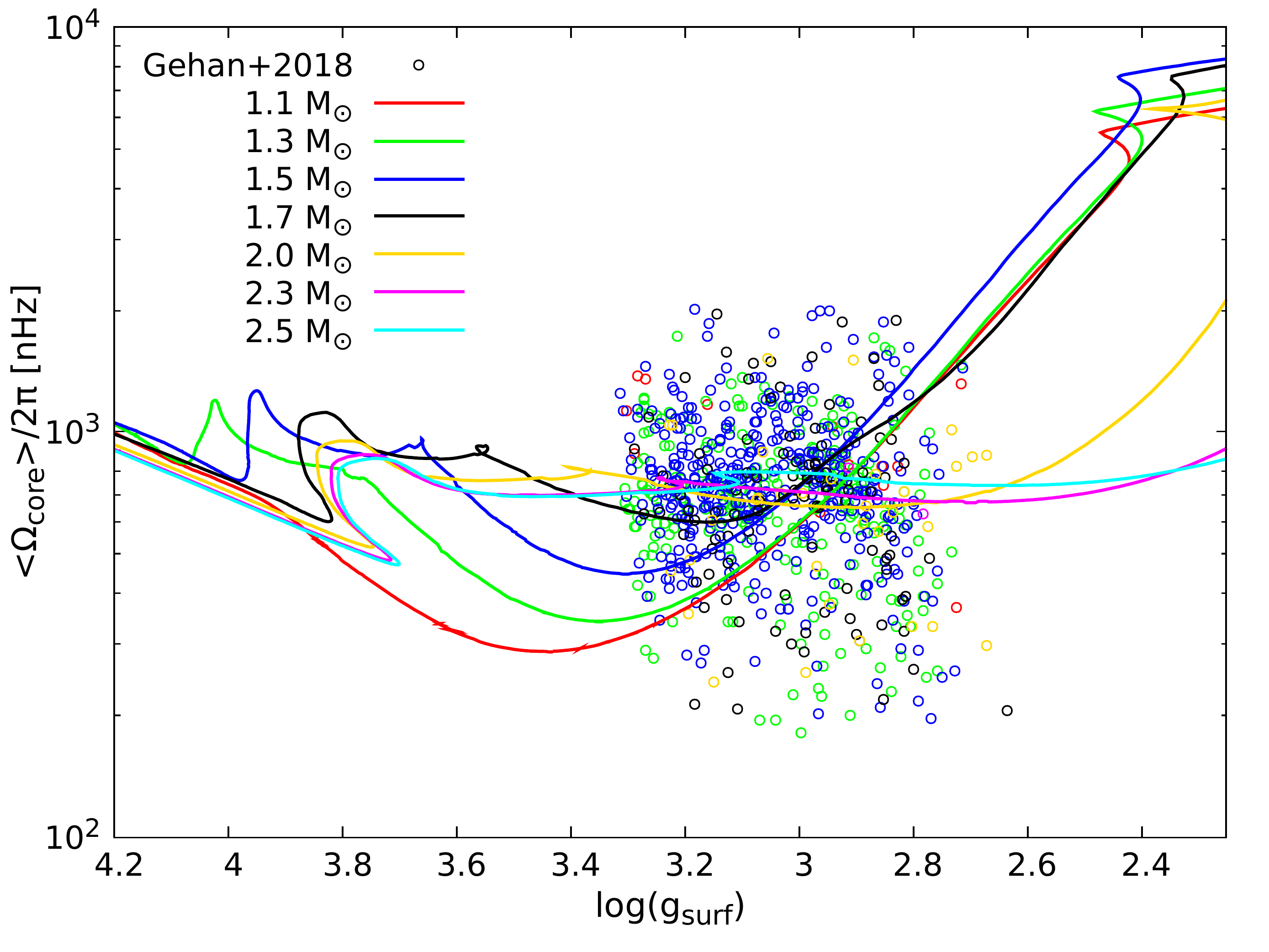}}
     \caption{Same as Fig. \ref{omegac_gsurf} but for models with the additional viscosity $\nu_{\rm add}$ needed to match the mean core rotation rate of red giants in the hydrogen shell-burning phase. The values of the additional viscosities for each model used in this figure are shown in Fig. \ref{nuadd_mass_rgb} and given in Table \ref{table_nuadd}.}
     \label{omegac_gsurf_nucal}
   \end{figure}

  In Fig. \ref{omegac_gsurf} we show a subset of our models with a constant additional viscosity of $\nu_{\rm add}=8 \times 10^{3}$ cm$^{2}$/s and the data set of red giants in the hydrogen shell-burning phase from \citet{gehan18}.
  We chose this value of $\nu_{\rm add}$ because it can reproduce the core rotation rate of most of the red giants for a stellar model with an initial mass of \mbox{M=1.1M$_{\odot}$}.
  The curves show the evolution of the mean core rotation rate for different initial masses, starting from the ZAMS and ending just above the RGB bump. This is seen at $\log g \sim$ 2.5 for stars with a mass lower than \mbox{M $\sim 2$M$_{\odot}$}.
  This figure illustrates two important points of this study: \textit{i)} the majority of stars group in a narrow band with \mbox{$\langle\Omega_{\rm core}/2\pi\rangle=$ 700 $\pm$ 100 nHz} and surface gravity in the range $\log g \sim$ 2.9 -- 3.3 (we refer to this band as the bulk of the data) and \textit{ii)} to correctly reproduce the bulk of the data, AM redistribution needs to increase its efficiency as mass increases in red giants close to the base of the RGB.

 \begin{figure}[htb!]
     \resizebox{\hsize}{!}{\includegraphics{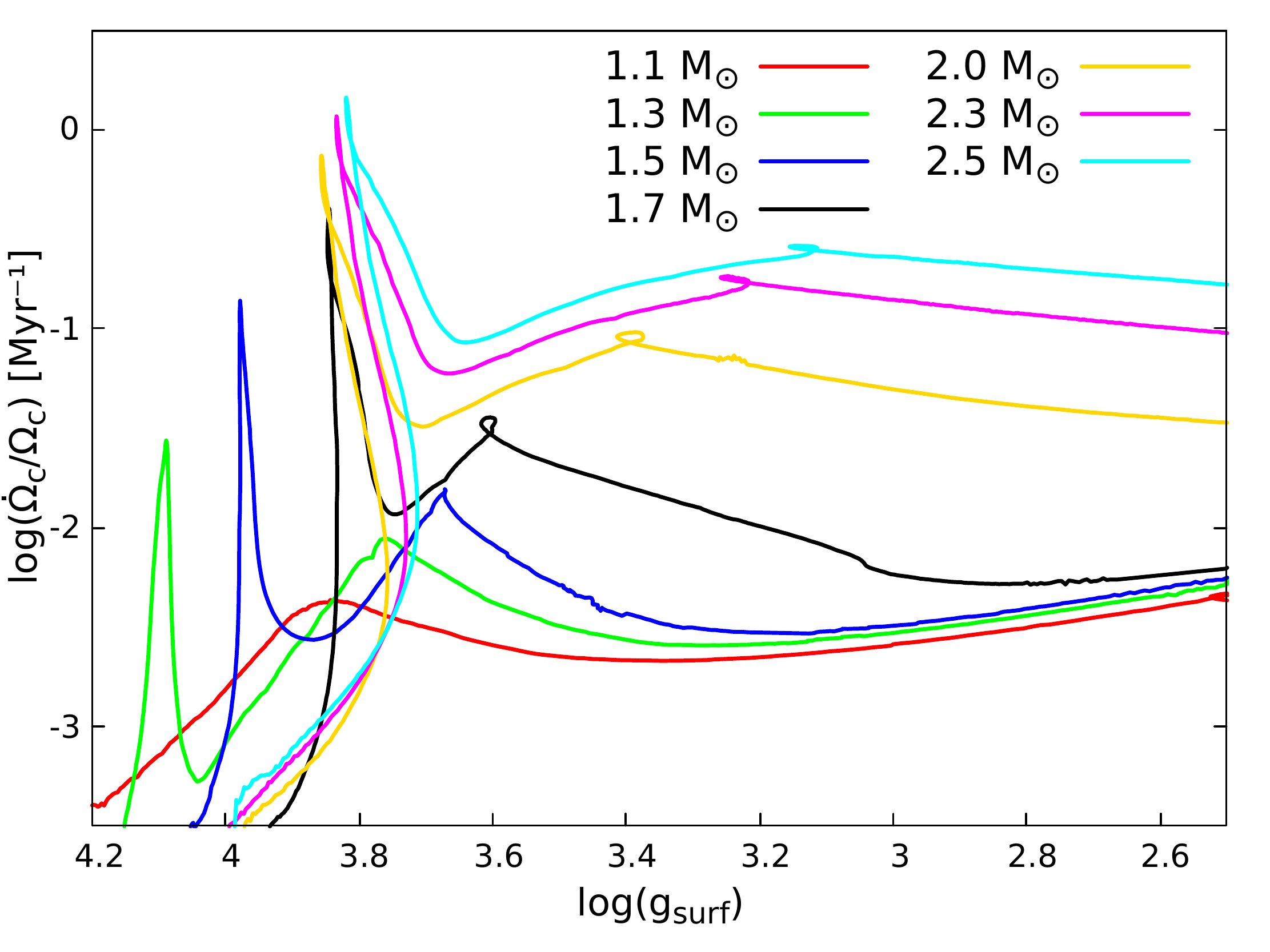}}
     \caption{Contraction rate of the core as a function of the surface gravity for models without including additional viscosity. Stars with mass lower than \mbox{M $\lesssim 2$ M$_{\odot}$} converge towards the same contraction rate at low surface gravity.}
     \label{omegadot_gsurf}
   \end{figure}

 \begin{figure}[htb!]
     \resizebox{\hsize}{!}{\includegraphics{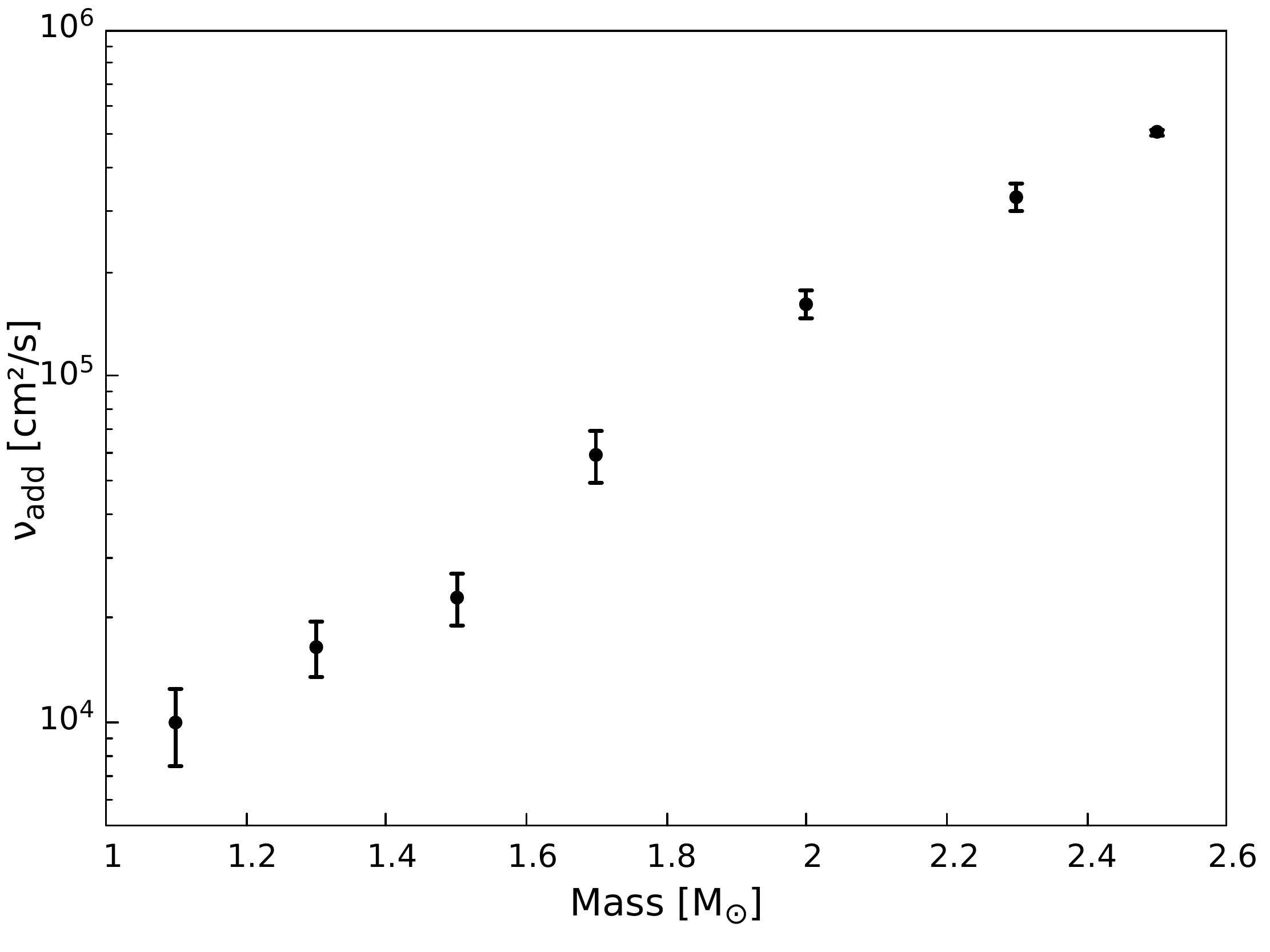}}
     \caption{Dependence of the additional viscosity (proxy for the efficiency of angular momentum redistribution) on the stellar mass in red giants in the hydrogen shell-burning phase. These values are estimated at a surface gravity of $\log g \sim 3.1$, or equivalently at a mixed-mode density of $\mathcal{N} \sim 7$. Including these values in stellar evolution models leads to core rotation rates in agreement with asteroseismic constraints. The values are given in Table \ref{table_nuadd}.}
     \label{nuadd_mass_rgb}
 \end{figure}

  \begin{figure}[htb!]
     \resizebox{\hsize}{!}{\includegraphics{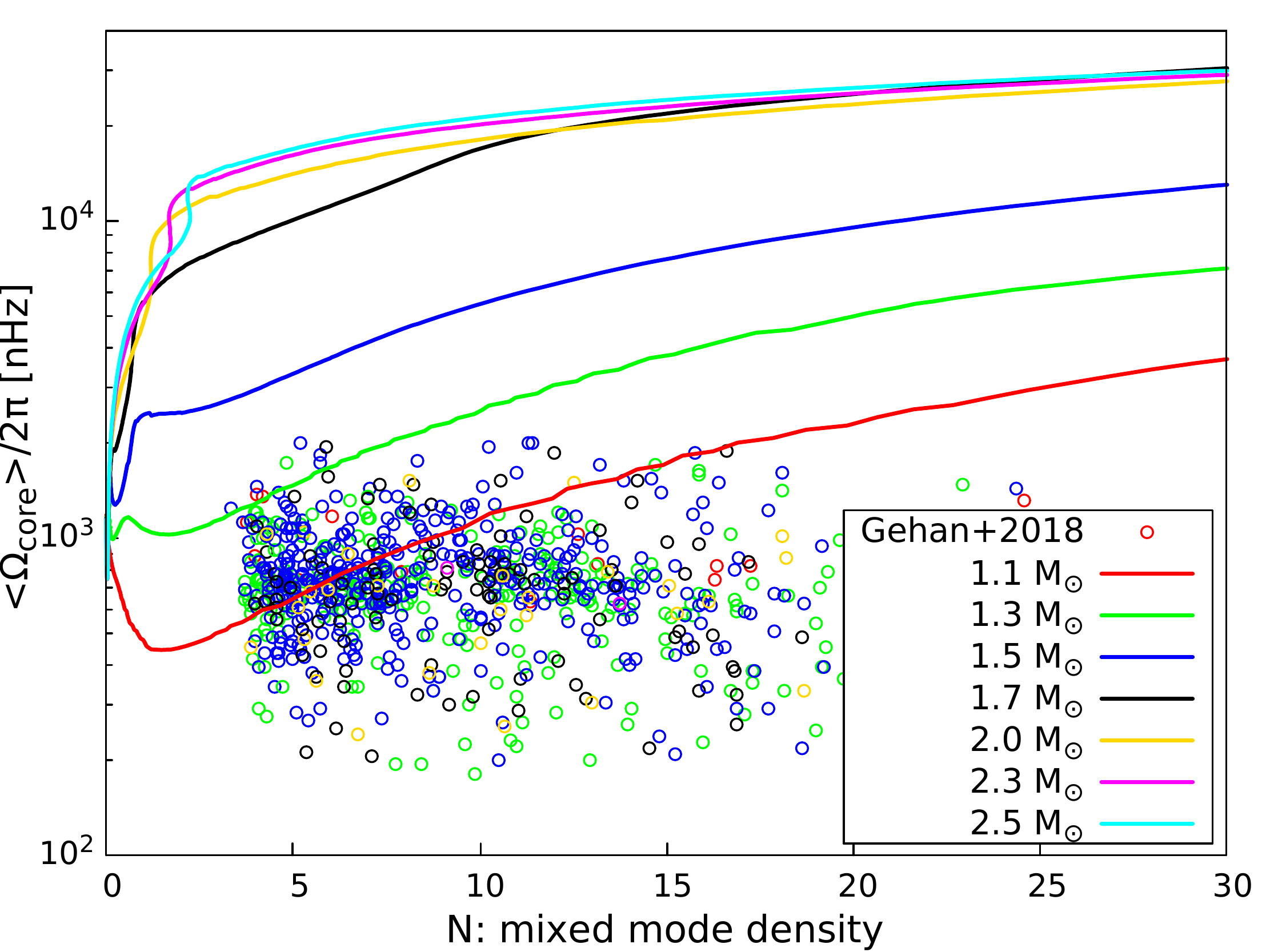}}
     \caption{Same as Fig. \ref{omegac_gsurf} but as a function of the mixed mode density $\mathcal{N}$.}
     \label{omegac_mixmod}
 \end{figure}

  To obtain a quantitative estimate of how the value of $\nu_{\rm add}$ varies with mass as well as on its uncertainty, we determined the value needed for each model in order to obtain a core rotation rate of $\langle \Omega_{\rm core}\rangle/2\pi = 700 \pm 100$ nHz at a surface gravity of $\log g = 3.1 \pm 0.2$ dex.
  For low-mass stars, the interval used in surface gravity represents the largest source of error in our estimates of $\nu_{\rm add}$. Indeed, in our models the core rotation rate increases significantly along evolution (see e.g. the red line in Fig. \ref{omegac_gsurf_nucal}). This behaviour requires the use of different $\nu_{\rm add}$ values along evolution to achieve a core rotation rate of $\langle\Omega_{\rm core}\rangle/2\pi \sim 700$ nHz, which are then used to estimate the uncertainty on $\nu_{\rm add}$ for each given stellar mass. On the other hand, the interval used in core rotation rate only results in a small uncertainty on $\nu_{\rm add}$ for low-mass stars. The situation is reversed for intermediate-mass stars (M $\gtrsim 2$ M$_{\odot}$). In this case the core rotation rate obtained in our models does not vary much along evolution in the region where measurements are available (see e.g. the cyan line in Fig. \ref{omegac_gsurf_nucal}). As a consequence, we can maintain a rate of $\langle\Omega_{\rm core}\rangle/2\pi \sim 700$ nHz with a roughly constant value of $\nu_{\rm add}$ along evolution on the lower RGB. The source of uncertainty thus here mainly comes from the interval used in core rotation rate. However, the evolution of the core rotation rate is very sensitive to the adopted $\nu_{\rm add}$ value, so that a small variations in $\nu_{\rm add}$ results in significantly different core rotation rates. Thus we end up with a small possible range of $\nu_{\rm add}$ values allowing to match the available measurements for intermediate-mass stars. This results in smaller uncertainties in $\nu_{\rm add}$ for intermediate-mass stars compared to low-mass stars.
  
 The resulting models with the values of $\nu_{\rm add}$ that we found are shown in Fig. \ref{omegac_gsurf_nucal}.
 In this figure, the core rotation rate of the low-mass models (M $\lesssim 2 M_{\odot}$) behave in a similar way once they start climbing the RGB (for $\log g \lesssim 3.5$).
   This occurs because the relative change in the core rotation rate ($\dot{\Omega}_{\rm c}/\Omega_{\rm c}$) of low-mass stars converges to roughly the same value towards the RGB (see Fig. \ref{omegadot_gsurf}).
   The first narrow peak located for example at $\log g \sim 4$ for the 1.5 M$_{\odot}$ model in Fig. \ref{omegadot_gsurf} occurs at the end of the main sequence and is a consequence of the rapid contraction in which the surface radius decreases and leads to the hook-like feature seen in the HR diagram.
   The second broader peak, occurring around $\log g \sim 3.6 -3.7$ for the 1.5 M$_{\odot}$ model in Fig. \ref{omegadot_gsurf}, is a second contraction of the central layers which occurs halfway through the subgiant phase \citep{eggenberger19a}.
   Since the additional viscosity sets the conditions for these models to have the same core rotation rate close to the base of the RGB, the increase of their core rotation rate seen in Fig. \ref{omegac_gsurf_nucal} is similar, leading to almost overlapping core rotation rates.
   The same cannot be said about the more massive models, with masses above \mbox{M $\gtrsim 2 M_{\odot}$} (see Fig. \ref{omegac_gsurf_nucal}), which have an increasing relative contraction rate $\dot{\Omega}_{\rm c}/\Omega_{\rm c}$ (or decreasing contraction timescale) with mass (see Fig. \ref{omegadot_gsurf}).

 The additional viscosity required to match the core rotation rate of the red giants is shown in Fig. \ref{nuadd_mass_rgb} as a function of the initial mass. We give the estimated values of $\nu_{\rm add}$ in Table \ref{table_nuadd}.
  We did the analysis using both the surface gravity and the mixed-mode density ($\mathcal{N}$) separately, and we found  consistent results.
  To estimate the uncertainty (error bars) in the estimations of $\nu_{\rm add}$ given in Table \ref{table_nuadd}, we consider the values of $\nu_{\rm add}$ that can reproduce a core rotation rate of $\langle\Omega_{\rm core}/2\pi\rangle = 700$ nHz at $\log g = 3.1$ allowing for an uncertainty of 100 nHz and 0.2 dex in core rotation rate and surface gravity, respectively.
 
    \begin{table}
      \caption[]{Values of the additional viscosities shown in Fig. \ref{nuadd_mass_rgb} and employed in Figs. \ref{omegac_gsurf_nucal} \& \ref{omegac_mixmod_nucal} to match the core rotation rate of red giants in the hydrogen shell-burning phase for different stellar masses.}
      \label{table_nuadd}
      $
         \begin{array}{p{0.5\linewidth}l}
            \hline
            \noalign{\smallskip}
            Mass [M$_{\odot}$]      &  \nu_{\rm add} [\rm{cm}^2/s] \\
            \noalign{\smallskip}
            \hline
            \noalign{\smallskip}
            1.1   &  (1.00\pm 0.25) \times 10^4  \\
            1.3   &  (1.65\pm 0.30) \times 10^4  \\
            1.5   &  (2.30\pm 0.40) \times 10^4  \\
            1.7   &  (5.90\pm 1.00) \times 10^4  \\
            2.0   &  (1.62\pm 0.15) \times 10^5  \\
            2.3   &  (3.30\pm 0.30) \times 10^5  \\
            2.5   &  (5.05\pm 0.10) \times 10^5  \\
            \noalign{\smallskip}
            \hline
         \end{array}
         $
   \end{table}

 \begin{figure}[!]
   \resizebox{\hsize}{!}{\includegraphics{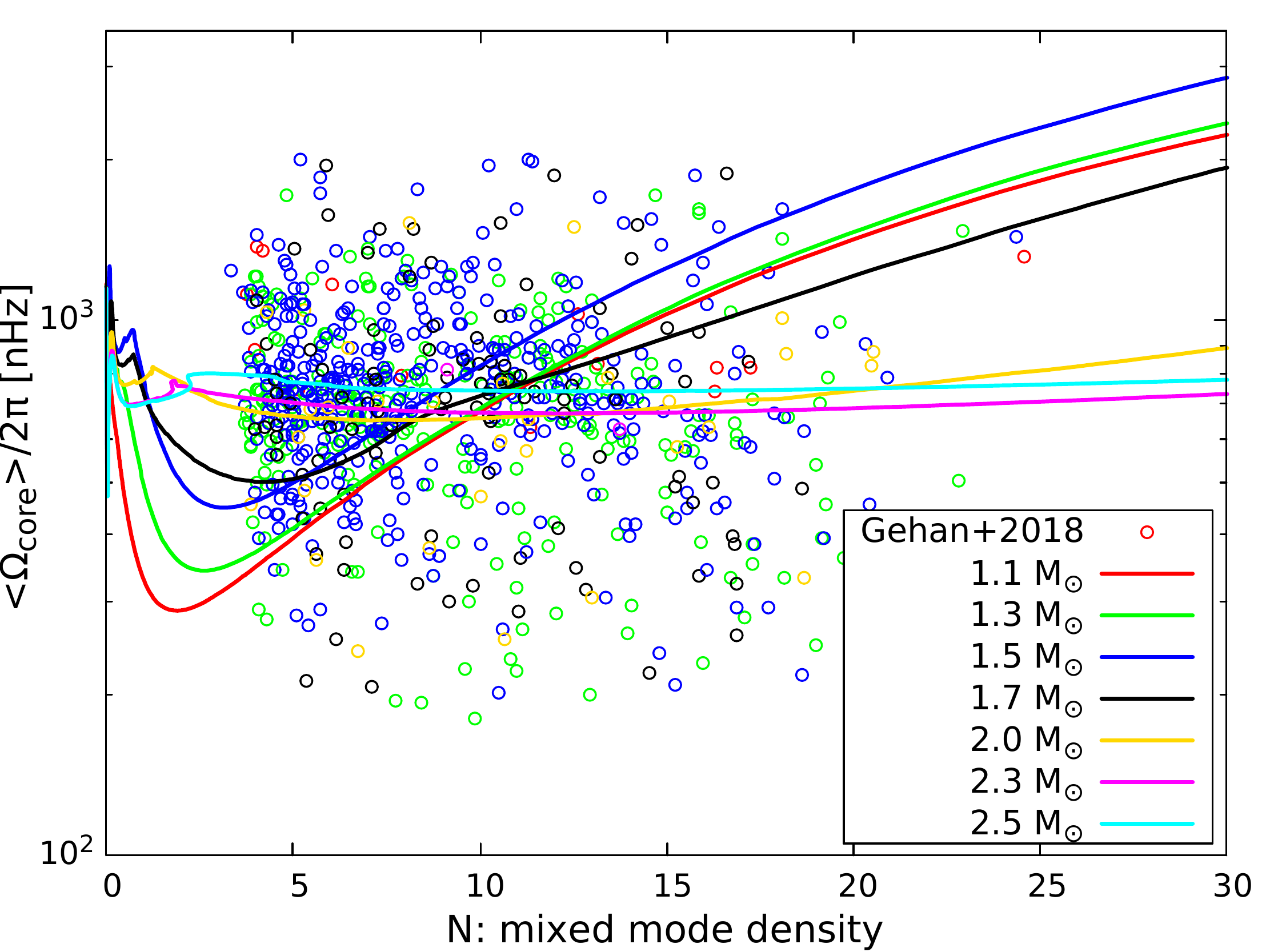}}
   \caption{Same as in Fig. \ref{omegac_gsurf_nucal} but as a function of the mixed-mode density.}
   \label{omegac_mixmod_nucal}
 \end{figure}

 \subsection{Dependence on the evolution along the red giant branch}
  \label{sect_rgbevol}
  To investigate if there is any dependence on the evolution along the RGB we resort to the mixed-mode density ($\mathcal{N}$).
  As shown by \citet{gehan18}, this quantity is an interesting proxy for the evolutionary stage of a star on the RGB, because it is mainly sensitive to the size of the helium-core in the red giant phase.
  We followed the same approach as in Sect. \ref{sect_rgbmass} to study the dependence of $\nu_{\rm add}$ on the evolution.
  The evolution of the core rotation rate as a function of the mixed-mode density for different initial masses is shown in Fig. \ref{omegac_mixmod}, for a subset of models with a constant additional viscosity of $\nu_{\rm add}=8 \times 10^{3}$ cm$^2$/s.
  And the corresponding models with the calibrated additional viscosities are shown in Fig. \ref{omegac_mixmod_nucal}.
  To quantify the evolution along the RGB based on the mixed-mode density we chose three representative values of the data bulk, at $\mathcal{N}=4, 9$ and 14.
 
  \begin{figure}[htb!]
    \resizebox{\hsize}{!}{\includegraphics{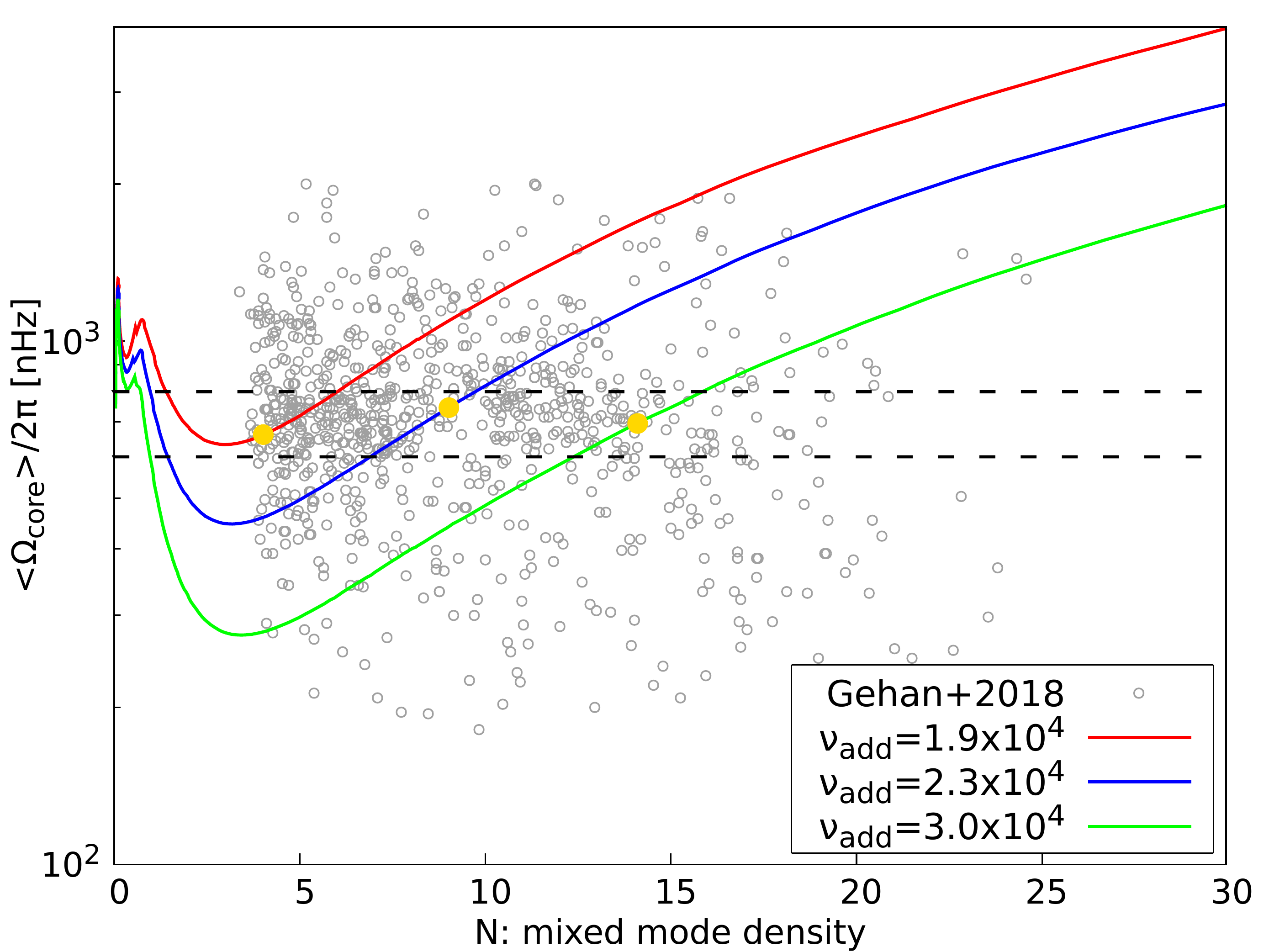}}
    \caption{Core rotation rate as a function of the mixed-mode density for models during the hydrogen-shell-burning phase. The efficiency of AM redistribution (parametrized by the additional viscosity $\nu_{\rm add}$) varies in order to reproduce the mean evolution of the core rotation rate as given by asteroseismic constraints, for a 1.5 M$_{\odot}$ model with an initial period of P$_{\rm ini}$=10 days at three different locations ($\mathcal{N}=4, 9$ and 14) shown with yellow dots on the evolutionary tracks. The dotted lines show the interval in which $\langle\Omega_{\rm core}\rangle/2\pi \in [600, 800]$ nHz.}
    \label{omegac_mixmod_nuevol}
  \end{figure}
  
\begin{figure}[htb!]
    \resizebox{\hsize}{!}{\includegraphics{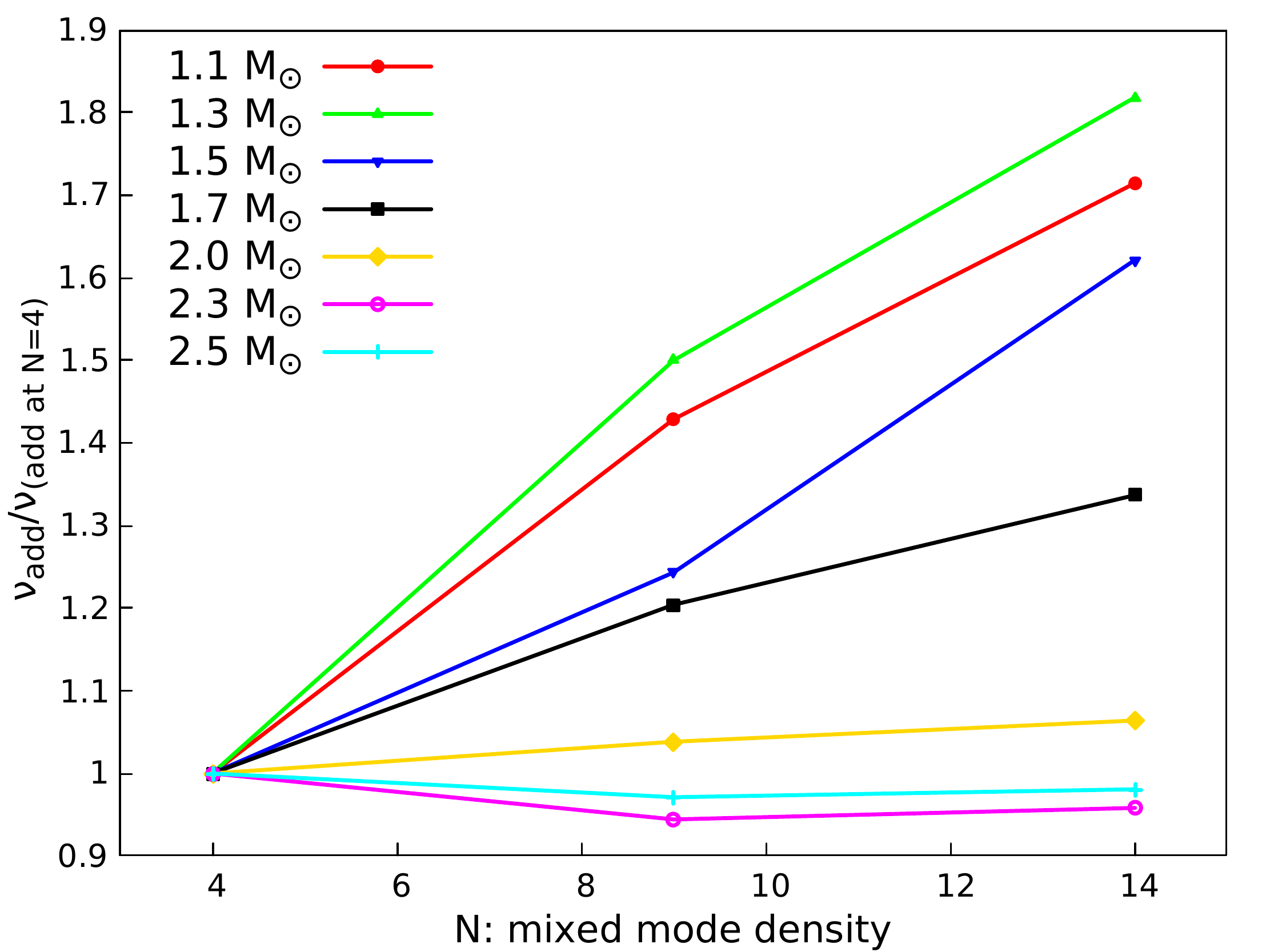}}
    \caption{Additional viscosity needed to satisfy the rotational constraints of red giants in the hydrogen shell-burning phase, normalized to its value at a mixed-mode density of $\mathcal{N}=4$, as a function of the mixed-mode density. The different curves correspond to different initial masses, as indicated in the legend.}
    \label{nuadd_rgb_mixmod}
  \end{figure}

  We illustrate this in Fig. \ref{omegac_mixmod_nuevol} for a 1.5 M$_{\odot}$ model in which we show the effect of changing the additional viscosity to match the core rotation rate at three different locations chosen in the data set.
  We show the values of $\nu_{\rm add}$ along the evolution on the RGB in Fig. \ref{nuadd_rgb_mixmod} for different initial masses.
  In that figure $\nu_{\rm add}$ is normalized to its value required to match the data bulk at $\mathcal{N}=4$ to disentangle the change along evolution from the mass dependence (see Fig. \ref{nuadd_mass_rgb}).

  For low-mass stars, the additional viscosity can increase up to a factor of two in the lower part of the RGB, while for stars with an initial mass of M $\gtrsim$ 2 M$_{\odot}$ the additional viscosity remains roughly constant (Fig. \ref{nuadd_rgb_mixmod}).
  Moreover, the variation of the additional viscosity along evolution tends to get weaker as the mass increases.
  Indeed, the normalized $\nu_{\rm add}$ values are lower as the mass increases, for each of the three $\mathcal{N}$ values considered, and even decreases slightly for the most massive ones.

  \subsection{Low-mass core-helium burning stars}
  \label{sect_redclump}
\begin{figure}[htb!]
     \resizebox{\hsize}{!}{\includegraphics{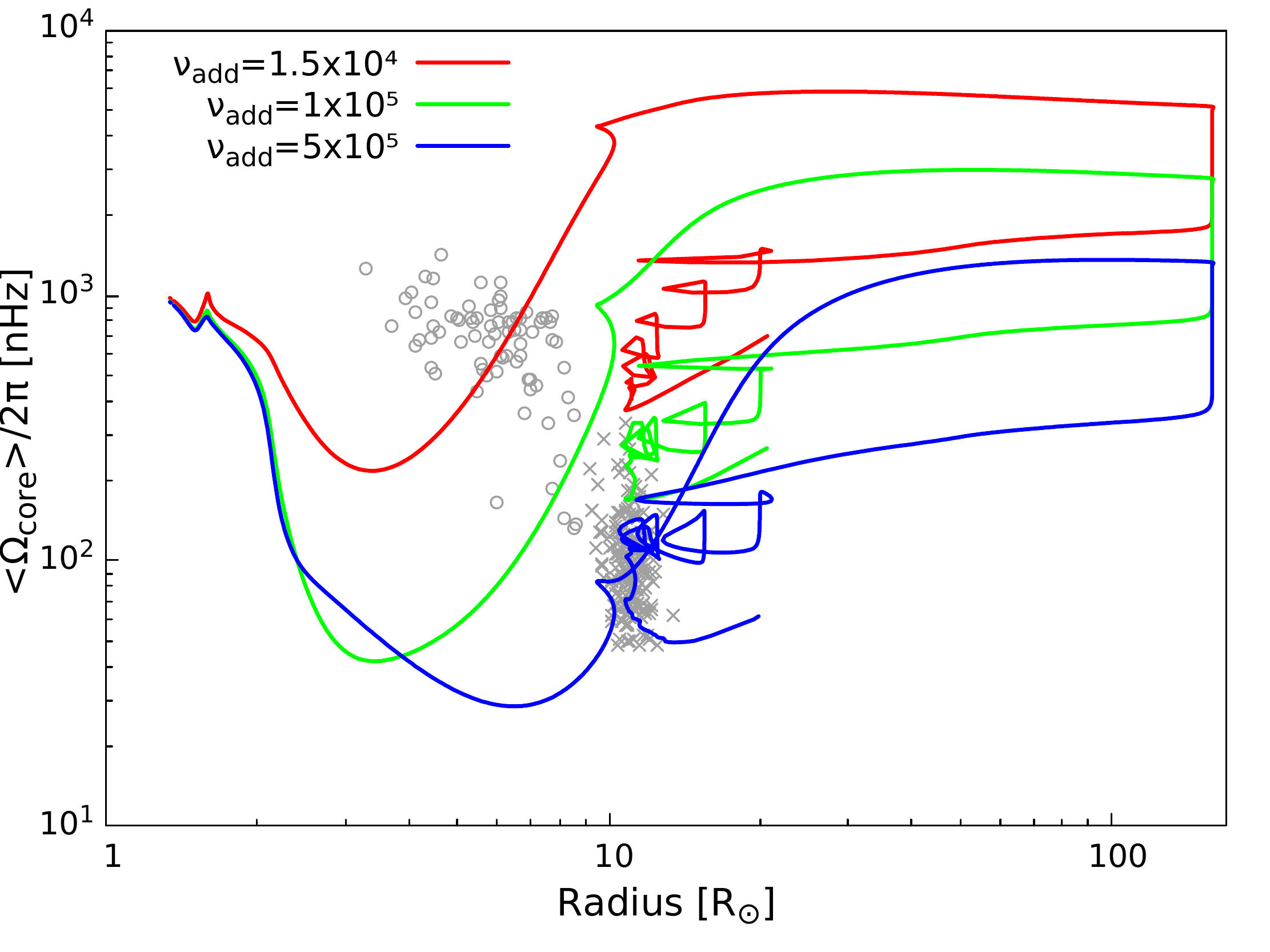}}
     \caption{Evolution of the core rotation rate as a function of the stellar radius. All models have an initial mass of 1.2 M$_{\odot}$ and an initial velocity of 5 km/s ($P_{\rm ini} =$ 12 days), and were computed with {\fontfamily{qcr}\selectfont MESA}. Different values for the additional viscosity are used in each model, as indicated in the figure in units of cm$^2$/s. The sharp drop in core rotation rate seen at R $\sim 150$ R$_{\odot}$ occurs because of the helium flash. The data correspond to red giants and red-clump stars from \citet{mosser12}. Only stars (data) with an estimated mass lower than M $\lesssim 2$ M$_{\odot}$ were included. Red-clump stars are grouped at radii R $\sim 10$ R$_{\odot}$ (crosses) while red giants in the hydrogen shell-burning phase are located above with a higher core rotation rate and lower radii (circles).
     }
     \label{redclump_grid_nu}
   \end{figure}

\begin{figure}[htb!]
\resizebox{\hsize}{!}{\includegraphics{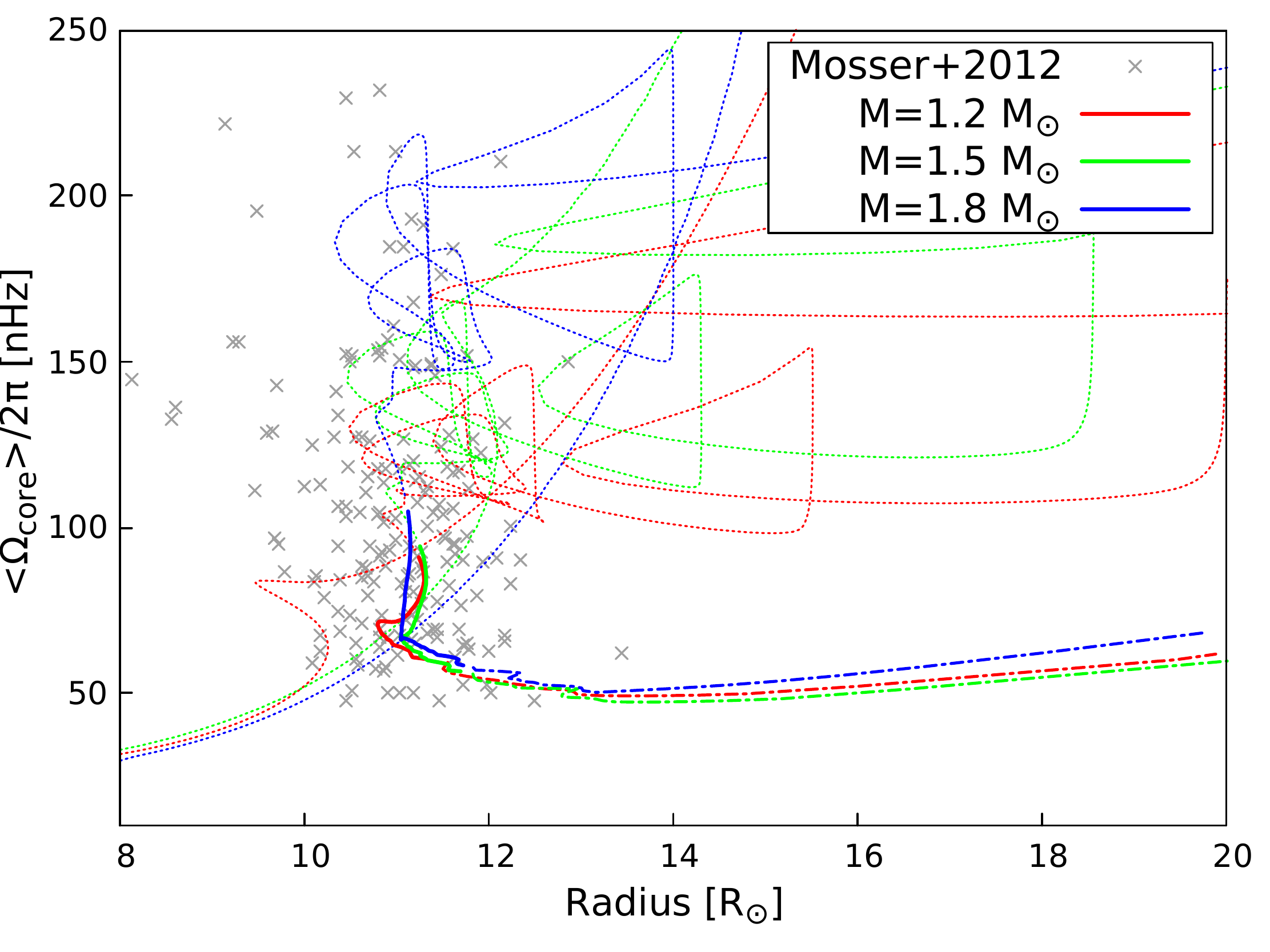}}
 \caption{Same as Fig. \ref{redclump_grid_nu} but for models with different initial masses, as indicated in the legend. We show only the region around the stable core-helium burning phase. All models have an additional viscosity of $\nu_{\rm add}=5 \times 10^{5}$ cm$^{2}$/s. The data points correspond to red-clump stars from \citet{mosser12}. \textit{Dotted lines}: previous to stable core-helium burning. \textit{Solid lines}: long stable core-helium burning phase (see text). \textit{Dash-dotted lines}: last fraction of stable core-helium burning phase. }
  \label{fig_redclump_mass}
   \end{figure}

   To study the efficiency of AM transport in red clump stars, we compute models until the end of the core-helium burning phase with initial masses of M$=1$ -- 2 M$_{\odot}$ using different additional viscosities, as outlined in Sect. \ref{sect_rgbmass}. The evolution of the core rotation rate as a function of the radius is shown in Figs. \ref{redclump_grid_nu} \& \ref{fig_redclump_mass}.
   In these figures we show the data of \citet{mosser12} for stars with a mass of M $\lesssim$ 2 M$_{\odot}$, to avoid including secondary clump stars which are intermediate-mass stars (2 $\lesssim$ M/M$_{\odot} \lesssim$ 8) in the core-helium burning phase \citep[e.g.][]{girardi98}.
   We verify that the additional viscosity must increase from the red giant phase towards the core-helium burning phase.
   This is because the additional viscosity calibrated for red giants is not sufficient for the red-clump stars.
   This is shown by the model with $\nu_{\rm add}=1.5 \times 10^4$ cm$^2$/s in Fig. \ref{redclump_grid_nu}, which can match the rotation rate of red giants but not that of core-helium burning stars.
   Considering the bounds on the core rotation rates from the data sets, we find that an additional viscosity of $\nu_{\rm add}= (3 \pm 2) \times 10^{5}$ cm$^{2}$/s is suitable to reproduce the core rotation rate of the red-clump stars (see Fig. \ref{redclump_grid_nu}).
   This value does not depend on the initial mass.

      The core slows-down the most during the first He-flash, when the degeneracy is lifted and the core expands.
      In this short phase, the core angular velocity decreases by $\sim 70 \%$. This is why in Fig. \ref{redclump_grid_nu} there is a sharp drop in the core rotation rate at a constant stellar radius of $R \sim 150 R_{\odot}$. 
      The relative decrease at constant radius in core rotation rate is independent of both the viscosity adopted and the initial mass, although the exact angular velocity of the core at the RGB tip depends on these two parameters.
      This phase is so rapid that the adopted value of the viscosity at the RGB tip does not make any difference in the subsequent evolution.
      The loops seen after the drop in $\langle\Omega_{\rm core}\rangle$, once the star contracts, are secondary flashes and last less than $\sim$ 4 Myr in total.
      Afterwards, the models enter into the stable core-helium burning phase (once the mass fraction of central helium drops below $\sim$ 0.9)  and spend more than $\sim 85$ \% of their burning time with a roughly constant radius of R $\sim 10$ -- 11 R$_{\odot}$. This corresponds to $\sim 100 $ Myr.

      The models with an additional viscosity of $\nu_{\rm add} \sim 10^5$ cm$^2/\rm{s}$, spend most of their core-helium burning time with a core rotation rate in agreement with the upper end of the data bulk for red-clump stars, which corresponds to $\langle\Omega_{\rm core}/2\pi\rangle \sim 150 $ nHz (see Fig. \ref{redclump_grid_nu}).
        Previous to this phase, they are in a rapid phase and thus we adopt $\nu_{\rm add} \sim 10^5$ cm$^2/\rm{s}$ as a lower limit for the additional vicosity.
        On the other hand, models with initial viscosities above $\nu_{\rm add} \sim 5 \times 10^5 \rm{ cm}^2/\rm{s}$ spend most of the time during the stable core-helium burning phase at lower core rotation rates than those of the data bulk (i.e. $\langle\Omega_{\rm core}/2\pi\rangle \lesssim 50 $ nHz). 
        Because of this we constrain the upper value of the additional viscosity to $\nu_{\rm add} \sim 5 \times 10^5$ cm$^2$/s.
        Therefore, an acceptable value of the additional viscosity for red-clump stars is $\nu_{\rm add} = (3 \pm 2) \times 10^5 $ cm$^2$/s.
      This value is weakly dependent on the stellar mass.
      We illustrate this in Fig. \ref{fig_redclump_mass} for models with different initial masses.
      The dotted lines show the region previous to the stable helium burning phase, solid lines show the region in which the model spends $\sim$ 85 \% of the duration in the core-helium burning phase while in the regions of dot-dashed lines it spends the remaining time until the end of the core-helium burning phase.
      This behaviour does not depend on the initial mass, thus we find that the inferred additional viscosity is weakly dependent on the stellar mass in this evolutionary phase. We note that it is difficult to reproduce the observed spread in radii in the red-clump with our models. This could be potentially accounted for with a distribution of initial metallicities, since we recall that in all our models a solar chemical composition is adopted.

   \section{Global behaviour of AM redistribution}
   \label{discussion}
   In the previous section, the efficiency of AM redistribution was determined for red giants in the hydrogen-shell and core-helium burning phases.
   Now we compare our results to those obtained in previous works for subgiants and secondary clump stars.
   In line with our work, the additional viscosity needed to satisfy the constraints on eight subgiants has been already obtained in previous works \citep{eggenberger19a,deheuvels20}.
   For two of these subgiants located close to the end of the main sequence, asteroseismic data are consistent with the fact that they are rotating as solid bodies. Because of this, \citet{deheuvels20} could only estimate a lower boundary on the additional viscosity, high enough to ensure the compatibility with asteroseismic constraints on core and surface rotation rates.

\begin{figure}[htb!]
     \resizebox{\hsize}{!}{\includegraphics{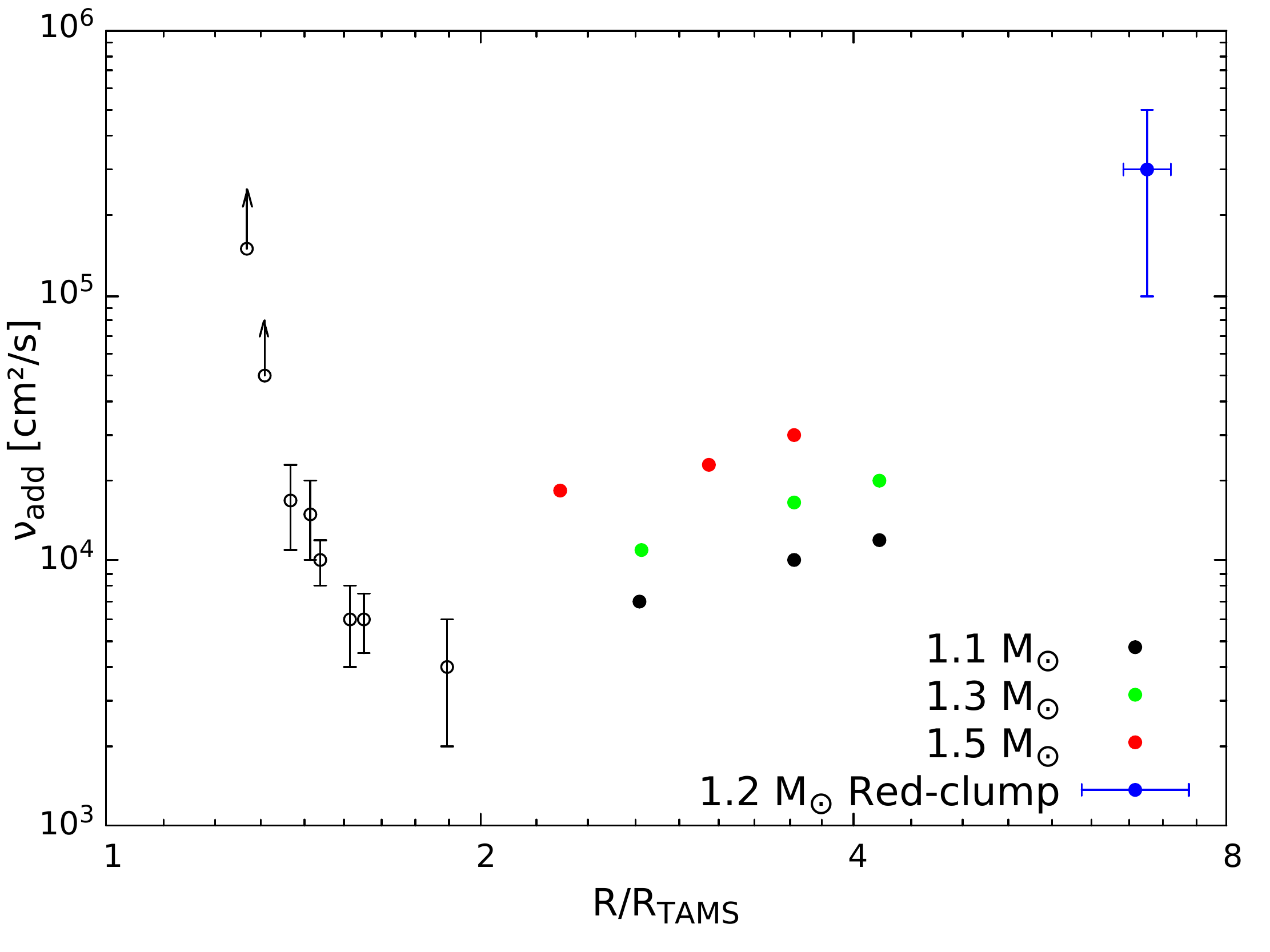}}
     \caption{Evolution of the efficiency of AM redistribution, parametrized by the additional viscosity $\nu_{\rm add}$, as a function of the radius normalized to the radius at the end of the MS. Black dots with $R/R_{\rm TAMS} < 2$ correspond to subgiants studied by \citet{deheuvels20} and \citet{eggenberger19a}. The other symbols correspond to red giants in the phase of hydrogen-shell-burning with different initial masses and finally to the estimation for red-clump stars based on a 1.2 M$_{\odot}$ model.}       
     \label{nuadd_evol}
   \end{figure}

   In Fig. \ref{nuadd_evol} we show the change in $\nu_{\rm add}$ with the evolution that we obtained for red giants compared to those inferred for subgiants.
   Following \citet{eggenberger19a}, we use the radius of the star normalized to its radius at the end of the main sequence as a proxy for the evolution.
   We note that the average mass of the subgiants is $M \sim 1.2 M_{\odot}$ and that the highest-mass star has an inferred mass of M=1.4 $M_{\odot}$.
   Since there is a mass-dependence for the red-giants that we studied, we show the evolution of $\nu_{\rm add}$ for different masses.
   It was already shown that the higher the stellar mass of subgiants, the more efficient the redistribution of AM \citep{eggenberger19b}.
   It also seems to be very efficient just after the main sequence, as inferred from the two young subgiants studied by \citet{deheuvels20}.
   Later in the subgiant phase its efficiency decreases until it reaches a minimum value close to the base of the RGB, to increase afterwards progressively as the star becomes a larger red giant.

   As for red-clump stars, we showed in Sect. \ref{sect_redclump} that an additional viscosity of $\nu_{\rm add}=(3 \pm 2) \times 10^{5} \rm{cm}^{2}/\rm{s}$ is needed to reproduce their core rotation rate.
   The horizontal error bars shown in Fig. \ref{nuadd_evol} come from the spread in radius of the data set.
   The fact that the efficiency needed to reproduce the core rotation rate of red giants is not enough for red-clump stars suggests that the efficiency must increase along the evolution on the red giant branch, coupling more strongly the core with the envelope.
   We do see this trend with evolution at least on the lower RGB (see Fig. \ref{nuadd_evol}).
   However, for red-clump stars we found an additional viscosity lower than what was obtained for secondary-clump stars by \citet{denhartogh19}.
   They found that the additional viscosity must have a value of  $\nu_{\rm add}=10^{7}$ cm$^{2}$/s, which is almost two orders of magnitude higher than for red-clump stars.
   Since our sample comprises only stars with a mass of M $\lesssim$ 2M$_{\odot}$, the higher value for secondary-clump stars possibly indicates a dependence on mass in the core-helium burning phase as well, or an indirect impact due to the structural and evolutionary differences between stars that burn helium in their core in a non-degenerate way (\mbox{M $\gtrsim$ 2 M$_{\odot}$}) compared to those that do not, and thus go through the helium flash phase (M $\lesssim$ 2 M$_{\odot}$).

   \subsection{Sensitivity to initial assumptions}
   \label{uncertainties}
   In our series of models we estimate the additional viscosity for red giants by adopting the same initial period chosen to correctly reproduce the surface rotation rates observed for subgiant stars.
   This assumption is simplistic since stars can be formed with different angular momenta.
   We thus discuss the impact of the initial rotation rate on the results obtained in the present study.

   We first note that the evolution of the core rotation rate in our models is similar when one adopts different initial periods with the calibrated values of the additional viscosity.
   We indeed checked that models with an initial period of either $P=2$ days or $P=50$ days lead to the same behaviour with evolution on the red giant branch, although the exact values of the additional viscosity may change depending on the initial period adopted.
   That is, the core of stars with a mass below 2 M$_{\odot}$ tends to spin up, while the rotation rate of more massive stars with a mass in the range $2 - 2.5$ M$_{\odot}$ tends to remain constant when we include an additional viscosity suitable to reproduce the rotational constraints.     
     As for the exact value of $\nu_{\rm add}$ and its sensitivity to the initial period, a 1.5 M$_{\odot}$ model would require for example an additional viscosity of $\nu_{\rm add}=5 \times 10^4 $ cm$^2$/s for an initial period of P$_{\rm ini}=2$ days or $\nu_{\rm add}=1 \times 10^4 $ cm$^2$/s for P$_{\rm ini}=50$ days to reproduce the data at a mixed-mode density of $\mathcal{N}=9$.
   Compared to the value of $\nu_{\rm add}=2.3 \times 10^4$ for a model with P$_{\rm ini}=10 $ days roughly a factor two on the estimation of the additional viscosity is needed for such a change in the initial period.
   Another constraint comes from the core-to-surface rotation ratios of red giants themselves \citep[e.g.][]{dim16,tri17,beck18,dimauro18,fellay21}.
   In our models the core-to-surface rotation ratio is weakly sensitive to the initial period adopted.
   Because of this a change in the initial period would not change the conclusions drawn in this work.
     
   In addition, we note that the surface rotation periods observed for MS stars with spectral type F-G are in the range $P \sim$ 5-25 days \citep[e.g.][]{mcquillan14,santos21,godoyrivera21}. Although the above cited works rely on the modulation of light-curves due to magnetic spots, that could lead to an underestimation in the number of slow rotators, it has been shown by \citet{masuda21} that the photometric samples do not lack slow rotators.
   The surface rotation periods obtained in our models during the MS are in good agreement with these observed values.

   We are aware that other uncertainties could affect the results obtained and the absolute values of the additional viscosity should be taken with caution.
   For example, a different initial metallicity can lead to either more compact or extended stars, as well as different evolutionary timescales, which can slightly change the exact values of the additional viscosities derived.
   However, we expect the trends with mass and evolutionary stage to remain unchanged.

   \subsection{Link with the physical nature of transport processes}
 This study allows us to characterize the efficiency of the internal AM transport in evolved stars, which is of prime interest to put constraints on the physical nature of the transport processes acting in stellar stable stratified regions. 
 The values derived for the needed additional viscosity and its variation with the mass and evolution can indeed serve as benchmark values for ongoing theoretical developments and numerical multi-dimensional simulations aiming at better understanding the dynamical processes in stellar interiors. 
 
     For instance, \citet{barker19,barker20} studied the role of the GSF \citep[][]{goldreich67,fricke68} instability in the transport of angular momentum via axisymmetric 3D hydro-dynamical simulations.
     Those authors suggest that an additional viscosity up to $\sim$ 10$^4$ cm$^2$/s can be attained by the GSF instability, and thus slow down the cores of red giants.
     Although it is a possible important contributor to the missing physical transport process, in particular for subgiants, the present study suggests that it would not be strong enough to slow down the cores of the more massive stars above $\sim 2$ M$_{\odot}$ , for which a much higher additional viscosity on the order of $\nu_{\rm add} \sim 10^5 - 10^6$ cm$^2$/s is required.
     
     AM transport by magnetic instabilities constitutes another promising candidate to explain the internal rotation of evolved stars. A first possibility is related to AM transport by the AMRI \citep{rud14}.
     Interestingly, this process could lead to an increase of the molecular viscosity by a factor of about 500 \citep[][]{rue15} that is then roughly compatible with the values found in the present study. It remains to be seen whether the trends with evolution and mass found here will be compatible with such a magnetic instability, which requires to take into account the strong inhibiting effects of chemical gradients located at the border of the helium core of red giants on this transport process.
     
     Previously, \citet{spada16} carried out a parametric study on the efficiency of the AM redistribution in red giants based on a power law of the ratio between the core and surface rotation rates ($\Omega_{\rm core}/\Omega_{\rm surf})^{\alpha}$ which can be related to the AMRI.
     Based on that relation with a power of $\alpha \sim 3$ they could reproduce the apparently decreasing trend in core rotation rate of red giants \citep{mosser12}.
     However, in the present study we benefit from a larger sample that indicates that the core rotation rate of red giants is roughly constant \citep{gehan18} rather than decreasing with evolution, which would be more favourable to a power of $\alpha \sim 2$ according to the results by \citet{spada16}.
       \citet{spada16} only computed models with a fixed mass of 1.25 M$_{\odot}$, which does not allow one to investigate the role of the stellar mass on the redistribution of AM.
       More work should then be done in this direction to compare our results with predictions of models accounting for AM transport by the AMRI.  

     Another possibility for AM transport by magnetic fields relies on the Tayler instability.
     As mentioned in the introduction, the Tayler instability is a potential important contributor to the AM transport in stellar interiors \citep[e.g.][]{spruit02,fuller19}, although in its present form it cannot reproduce simultaneously the core rotation rates of subgiants and red giants \citep{eggenberger19c,den20}.
     The results of the present study could help improve the modelling of these processes of both the AMRI and the Tayler instabilities.
     
     Other physical processes at play in this kind of stars like (but not limited to) wave flux of angular momentum by internal gravity waves \citep[e.g.][]{pincon16,pincon17} or mixed modes \citep{belkacem15} should be tested in detail in advanced phases with stellar evolutionary calculations to shed light on the physical mechanism redistributing angular momentum in stellar interiors.

   \section{Conclusions}
   \label{conclusion}
   In this work we studied the efficiency with which AM must be redistributed in stellar interiors to match the constraints on the core rotation rate of red giants in the hydrogen-shell and core-helium burning phases.
   For this, we computed stellar evolution models including an additional viscosity with the {\fontfamily{qcr}\selectfont GENEC} and {\fontfamily{qcr}\selectfont MESA} stellar evolution codes.
   With this approach, we infer a mean value of the diffusion coefficient that transports angular momentum.
   We compared our models to a data set of roughly nine hundred red giants obtained by \citet{mosser12} and \citet{gehan18}, including red giants in the hydrogen shell-burning phase close to the base of the red giant branch and more evolved low-mass stars in the core-helium burning phase.
   We did not aim to model individually each star, but to investigate the behaviour of the efficiency needed to match the constraints on their core rotation rates.
   We then compare our results with previous works done for subgiants and secondary-clump stars to have a picture of how the efficiency of the physical process should vary along evolutionary timescales for different stellar masses.
   Our main conclusions are listed below:
   \begin{enumerate}
   \item The efficiency of internal AM transport must decrease with evolution in the subgiant phase and become stronger once the star evolves towards higher luminosities on the red giant branch.
   \item For red giants in the hydrogen-shell burning phase, the core rotation rate of intermediate-mass stars can be reproduced with a constant AM redistribution efficiency, while for lower mass stars the efficiency must increase as they evolve.
   \item For red giants in the hydrogen-shell burning phase, AM redistribution needs to be more efficient for higher-mass stars.
   \item In red-clump stars (low-mass stars burning He in their core), AM must be redistributed between one and two orders of magnitude more efficiently than for red giants close to the base of the red giant branch and is independent of the mass. Also, the efficiency needed for red-clump stars is lower than what was already found for secondary-clump (intermediate-mass stars burning He in their core) stars.
   \end{enumerate}

  All these trends about AM redistribution in evolved stars constitute key constraints that any physical candidate transport mechanism should be able to reproduce.

\begin{acknowledgements}
  F.D.M., P.E., G.M. and S.J.A.J.S. acknowledge funding by the  European Research Council (ERC) under the European Union’s Horizon 2020 research and innovation program (grant agreement No. 83925, project STAREX).
  G.B acknowledges fundings from the SNF AMBIZIONE grant No. 185805 
(Seismic inversions and modelling of transport processes in stars).
\end{acknowledgements}

%
%
\bibliographystyle{aa}
\bibliography{const_nu}

\end{document}